\documentclass{ws-ijgmmp}
\usepackage{notoccite}
\usepackage[usenames,dvipsnames,svgnames]{xcolor}
\usepackage[colorlinks=true,
      linkcolor=red,
      urlcolor=gray,
      citecolor=blue]{hyperref}
\usepackage[]{cite}
\usepackage{float}
\begin{document}

\markboth{Joseph Ntahompagaze, Amare Abebe,Manasse Mbonye}
{On $f(R)$ gravity in scalar-tensor theories}

%
\catchline{}{}{}{}{}
%

\title{On $f(R)$ gravity in scalar-tensor theories}

\author{Joseph Ntahompagaze }

\address{Astronomy and Astrophysics Division, Entoto Observatory and Research Center,\\
Addis Ababa, Ethiopia,\\
Department of Physics, College of Science and Technology, University of Rwanda, \\
 Kigali, Rwanda,\\
ntahompagazej@gmail.com}

\author{Amare Abebe}

\address{Department of Physics, North-West University, \\
Mafikeng, South Africa\\
Astronomy and Astrophysics Division, Entoto Observatory and Research Center,\\
Addis Ababa, Ethiopia,\\
amare.abbebe@gmail.com}

\author{Manasse Mbonye}

\address{Department of Physics, College of Science and Technology, University of Rwanda, \\
 Kigali, Rwanda,\\
mmbonye@gmail.com}

\maketitle

\begin{history}
\end{history}

\begin{abstract}
We study $f(R)$ gravity models in the language of scalar-tensor theories. The correspondence between $f(R)$ gravity and scalar-tensor theories 
is revisited since $f(R)$ gravity is a subclass of Brans-Dicke models, with a vanishing coupling
constant ($\omega=0$). In this treatment, four $f(R)$ toy  models are used to analyze the early-universe cosmology, when the scalar field $\phi$
dominates over standard matter. We have obtained solutions to the Klein-Gordon equation for those models. 
It is found that for the first model $\left(f(R)=\beta R^{n}\right)$, as time increases the scalar field decreases and decays asymptotically.
For the second model $\left(f(R)=\alpha R+\beta R^{n}\right)$ it was found that the function $\phi(t)$ crosses  the $t$-axis at different values for different values of
$\beta$. For the third model $\left(f(R)=R-\frac{\nu^{4}}{R}\right)$, when the value of $\nu$ is small 
the potential $V(\phi)$  behaves like the standard inflationary potential. For the fourth model 
$\left(f(R)=R-(1-m)\nu^{2}\Big(\frac{R}{\nu^{2}}\Big)^{m}-2\Lambda\right)$, we show that there is a transition between $1.5<m<1.55$. The behavior of the potentials
with $m<1.5$ is totally different from those with $m>1.55$. The slow-roll approximation is applied to each of the four $f(R)$ models 
and we obtain the respective expressions for the  spectral index $n_{s}$ and  the tensor-to-scalar ratio $r$.  
\end{abstract}

\keywords{$f(R)$ gravity --- scalar-tensor --- scalar field ---cosmology.}
\ccode{PACS numbers: 04.50.Kd, 98.80.-k, 95.36.+x, 98.80.Cq; MSC numbers: 83Dxx, 83Fxx}
 
  \section{Introduction} 

The advancement in modern observations in astrophysics and cosmology is increasing very fast. This makes the underlying theories 
to be left behind observations. Different theories have been constructed and others are under construction to predict and explain the 
observations in cosmology. Some of them are quantum gravity, string theory, scalar-tensor theory, etc. Those theories are known as
modified gravity theories since they modify General Relativity (GR) constructed by Einstein (see extensive review in \cite{scalar1}).
The scalar-tensor (ST) theory is a higher-order theory, where degrees of freedom to explain different scenarios are present. Some of those 
degrees  of freedom are scalar field, the coupling constant and cosmological constant as well \cite{scalar2}. The Brans-Dicke (BD) theory 
is one of the special classes of the ST theory, where the coupling parameter $\omega(\phi)$ is supposed to be independent of
the scalar field $\phi$. It is considered to be constant and hence the name `coupling constant' $\omega$. The BD theory has attracted 
researchers to test its predictions on solar-system scales. The  most interesting outcome is that the coupling constant has to be large 
to approximate the solar-system predictions \cite{dicke3}.\\
Also due to the push from the observational cosmology, the ST theory has been succeeding in approximating some observable 
parameters such as the spectral index $n_{s}$ and tensor-to-scalar ratio $r$ in good accuracy (see Ref. \cite{slow3,slow4} ).  \\ 
In this paper, we present $f(R)$ gravity theory as a subclass of ST theory. In this analysis, four $f(R)$ models have been taken into account.
We have derived the respective potential $V(\phi)$ for each $f(R)$ model. Then, the Klein-Gordon equation (KGE) has been obtained for each 
model and the corresponding solutions have been calculated. \\
The paper is organized as follows. In the next section we review the field equations. Then, we review $f(R)$ theory in ST language in Section \ref{frst}. In Section \ref{sla} the slow-roll approximation of the four $f(R)$ models have been done. The last section is devoted for the conclusion.

The adopted spacetime signature is $(-+++)$ and unless stated otherwise, we have used the convention $8\pi G=c=1$, where $G$ is the gravitational constant and $c$ is the speed of light.

  \section{Field Equations}

The Einstein Field Equations (EFEs) can be derived from the Riemann geometry (the Bianchi identities, Ricci tensor, Ricci scalar and Einstein tensor) 
or from Hilbert-Einstein action, given by 
\begin{equation}
I_{GR}=\frac{1}{2\kappa}\int d^{4}x \sqrt{-g}[R+\mathcal{L}_{m}], \label{action1}
\end{equation}
where $\mathcal{L}_{m}$ is the matter Lagrangian, and  $\kappa\equiv 8\pi G/c^{4}=1$ by virtue of the convention described above.

For $f(R)$ gravity, the action in Eq. \eqref{action1} has to be modified, where instead of having the Ricci scalar
$R$, we have the function $f(R)$ (see Ref. \cite{amare1,7}). This function depends on the Ricci scalar $R$.
So the action reads
\begin{equation}
I_{f(R)}=\frac{1}{2\kappa}\int d^{4}x\sqrt{-g}[f(R)+\mathcal{L}_{m}].\label{6}
\end{equation} 
In $f(R)$ gravity theory, one can vary the action with respect to the metric only (results in metric formalism), or with respect to the other parameters (this results in Palatini formalism and metric-affine formalism depending on the variation type). Here  we will be interested in metric-formalism.\\
The action in Eq. \eqref{6} produces the field equation (see Ref. \cite{amare1,7,amare3}):
\begin{equation}
G_{\mu\nu}=\frac{1}{f'}\big[T^{m}_{\mu\nu}+\frac{1}{2}g_{\mu\nu}(f-Rf')+\nabla_{\nu}\nabla_{\mu}f'-g_{\mu\nu}\nabla_{\sigma}\nabla^{\sigma}f'\big],\label{eqfr0}
\end{equation}
where $f=f(R)$,$f'=\frac{df}{dR}$ and $T^{m}_{\mu\nu}=-\frac{2}{\sqrt{-g}}\frac{\delta (\sqrt{-g}\mathcal{L}_{m})}{\delta g^{\mu\nu}}$ is the energy-momentum tensor (EMT) of standard matter. 
If a 1+3 decomposition is made such that 
\begin{equation}
g_{ab}=h_{ab}-u_{a}u_{b}, 
\end{equation}
where $h_{ab}$ is a projected tensor and $u^{a}$ is a 4-vector field perpendicular to the hypersurfaces of constant curvature of spacetime (see Ref. \cite{3,amare1,amare2}), 
then we can have energy momentum tensor EMT for the matter-curvature composition:
\begin{equation}
T_{ab}=T^{m}_{ab}+T^{R}_{ab}=\mu u_{a}u_{b}+q_{a}u_{b}+u_{a}q_{b}+ph_{ab}+\pi_{ab}\;,\label{13a}
\end{equation}
where $\mu =\tilde{\mu}_{m}+\mu_{R}$, $p=\tilde{p}_{m}+p_{R}$, $q_{a}=\tilde{q}^{m}_{a}+q^{R}_{a}$ and $\pi_{ab}=\tilde{\pi}^{m}_{ab}+\pi^{R}_{ab}$, with $\tilde{\mu}_{m}=\frac{\mu_{m}}{f'}$, $\tilde{p}_{m}=\frac{p_{m}}{f'}$, $\tilde{q}^{m}_{a}=\frac{q^{m}_{a}}{f'}$ and $\tilde{\pi}^{m}_{ab}=\frac{\pi^{m}_{ab}}{f'}$. Here $T^{m}_{ab}$ stands for the total EMT of standard matter fluids and $T^{R}_{ab}$ is the EMT of the curvature fluid.\\ 
The trace of Eq. \eqref{13a} is 
\begin{equation}
T=T^{a}_{a}=T^{m}+T^{R}=3p-\mu=(3\tilde{p}_{m}-\tilde{\mu}_{m})+(3p_{R}-\mu_{R}),\label{13b}
\end{equation}
where $T^{m}=3\tilde{p}_{m}-\tilde{\mu}_{m}$ and $T^{R}=3p_{R}-\mu_{R}$.
In this work, we have considered Friedmann-Lema\^itre-Robertson-Walker (FLRW) spacetime universe.
Therefore quantities like energy density and isotropic pressure for perfect-curvature fluid in $f(R)$ theories are given 
in Ref. \cite{amare1,amare3} as :  

\begin{equation}
\mu_{R}=\frac{1}{f'}\big[\frac{1}{2}(Rf'-f)-\Theta f'' \dot{R}\big],\label{14a}
\end{equation}
\begin{equation}
p_{R}= \frac{1}{f'}\big[\frac{1}{2}(f-Rf')+f''\ddot{R}+f'''\dot{R}^{2}+\frac{2}{3}\Theta f'' \dot{R}\big],\label{14b}
\end{equation}
and 
\begin{equation}
q^{R}_{a}=\pi^{R}_{ab}=0,
\end{equation}
where $\Theta$ is volume rate of expansion of the fluid.
Using the equations \eqref{14a} and \eqref{14b} in Eq. \eqref{13b}, we have
\begin{equation}
T=T^{a}_{a}=3\tilde{p}_{m}-\tilde{\mu}_{m}+\frac{1}{f'}\big[2(f-Rf')+3(\Theta f''\dot{R}+f''\ddot{R}+f'''\dot{R}^{2})\big],\label{T1a}
\end{equation}
where the trace of the curvature EMT is  
\begin{equation}
T^{R}=\frac{1}{f'}\big[2(f-Rf')+3(\Theta f''\dot{R}+f''\ddot{R}+f'''\dot{R}^{2})\big].\label{T1b}
\end{equation}

\section{$f(R)$ Gravity in the Scalar-Tensor Language}\label{frst}

ST theories have been a point of interest in cosmology for quite some time now \cite{ scalar1,scalar2,7}. $f(R)$ gravitational models  have been shown to be 
a sub-class of the ST theory \cite{scalar3}.
A clear illustration of how $f(R)$ theories are classified as a subclass of ST theory is in the BD theory for 
the case of the coupling constant $\omega=0$ \cite{7}. The ST theory has the general action, see Ref. \cite{scalar1}:
\begin{equation}
 I_{ST}=\int d^{4}x \sqrt{-g}\big[f(\phi)R-g(\phi)\nabla_{\mu}\phi\nabla^{\mu}\phi-2\Lambda(\phi) 
        +\mathcal{L}_{m}(\Psi,g_{\mu\nu})\big],\label{stt0}
\end{equation} 
where $f,g$ and $\Lambda$ are arbitrary functions of the scalar field $\phi$ and $\mathcal{L}_{m}$ is the Lagrangian density of 
the matter field $\Psi$. If a redefinition is made such that $f(\phi)=\phi$ and by introducing
the coupling parameter $\omega(\phi)$ instead of $g(\phi)$, we have (see Ref.\cite{scalar1}): 
\begin{equation}
I_{ST}=\int d^{4}x \sqrt{-g}\big[\phi R-\frac{\omega(\phi)}{\phi}\nabla_{\mu}\phi\nabla^{\mu}\phi-2\Lambda(\phi) +\mathcal{L}_{m}(\Psi,g_{\mu\nu})\big].\label{stt1}
\end{equation}
The variation of the action in Eq. \eqref{stt1} with respect to the metric $g_{\mu\nu}$ gives the field equations as
\begin{equation}
\begin{split}
\phi G_{\mu\nu}+\big[\square \phi +\frac{1}{2}\frac{\omega}{\phi}(\nabla \phi)^{2}+\Lambda \big]g_{\mu\nu}-\nabla_{\mu}\nabla_{\nu}\phi
 -\frac{\omega}{\phi}\nabla_{\mu}\phi\nabla_{\nu}\phi=8\pi T_{\mu\nu}. \label{stt3}
\end{split}
\end{equation}
The trace of Eq. \eqref{stt3} together with the elimination of $R$ gives
\begin{equation}
(2\omega+3)\square \phi +\omega'(\nabla \phi)^{2}+4\Lambda-2\phi \Lambda'=8\pi T,
\end{equation}
where primes here denote partial differentiation with respect to $\phi$.\\
The action in Brans-Dicke theory can  be recovered by setting $\omega$ to be a constant and making $\Lambda(\phi)=0$ in Eq. \eqref{stt1}, see Ref. \cite{scalar1,7}.
Thus we have
\begin{equation}
I_{BD}=\int d^{4}x \sqrt{-g}\big[\phi R-\frac{\omega}{\phi}\nabla_{\mu}\phi\nabla^{\mu}\phi+\mathcal{L}_{m}(\Psi,g_{\mu\nu})\big].\label{stt2}
\end{equation} 
Let us consider the action that represents $f(R)$ gravity given as
\begin{equation}
I=\frac{1}{2\kappa}\int d^{4}x\sqrt{-g}[f(R)+\mathcal{L}_{m}].
\end{equation}
The action in ST theory has the form (see Ref. \cite{amare1,scalar5}):
\begin{equation}
I_{f(\phi)}=\frac{1}{2\kappa}\int d^{4}\sqrt{-g}\left[f(\phi (R))+\mathcal{L}_{m}\right],\label{frstt1}
\end{equation}
where $f(\phi(R))$ is the function of $\phi(R)$ and we consider the scalar field $\phi$ to be
\begin{equation}
\phi=f'-1.
\end{equation}
Here the prime indicates differentiation with respect to $R$ and  the scalar field $\phi$ should be invertible \cite{scalar1,7,scalar3}.
Thus if we compare this action to that in Brans-Dicke theory (see Eq. \eqref{stt2}) for the case of vanishing coupling constant $\omega=0$, we can say 
that  $f(R)$ theory is a special case of the ST theory. Note also that this is the Jordan frame representation and the action has a different
form when the Einstein frame is used. One can use the Palatini approach to show that $f(R)$ is a sub-class of scalar ST theory but in that context, the coupling 
constant is considered to be $\omega=-\frac{3}{2}$ (see more detail in Ref. \cite{scalar1,scalar4}).
The field equations from the action in Eq. \eqref{frstt1} are given in Ref. \cite{7} as:
\begin{equation}
 G_{ab}=\frac{\kappa}{\phi+1}T^{m}_{ab}+\frac{1}{(\phi+1)}\left[\frac{1}{2} g_{ab}\left(f-(\phi+1)R\right)
 +\nabla_{a}\nabla_{b}\phi-g_{ab}\square \phi \right],\label{frstt2a}
\end{equation}
where $\square =\nabla_{c}\nabla^{c}$ is the covariant D'Alembert operator. The EMT for the scalar field is given as
\begin{equation}
T^{\phi}_{ab}=\frac{1}{(\phi+1)}\left[\frac{1}{2} g_{ab}\big(f-(\phi+1)R\big)+\nabla_{a}\nabla_{b}\phi-g_{ab}\square \phi\right].
\end{equation}
The scalar filed $\phi$ obeys the Klein-Gordon Eq. (see Ref. \cite{amare1, scalar5}):
\begin{equation}
\square \phi -\frac{1}{3}\big(2f-(\phi+1)R+T^{m}\big)=0, \label{KG}
\end{equation}
where $T^{m}$ is the trace of the matter EMT. \\
One can consider at this stage the Friedmann and Raychaudhuri equations \cite{amare1,amare3}:
\begin{equation}
\Theta^{2}=3(\tilde{\mu}_{m}+\mu_{R})-\frac{9K}{a^{2}},\label{fried}
\end{equation}
and 
\begin{equation}
\dot{\Theta}+\frac{1}{3}\Theta^{2}+\frac{1+3\omega}{2}\tilde{\mu}_{m}+\frac{1}{2}(\mu_{R}+3p_{R})=0,\label{duri}
\end{equation}
respectively, where $K$ stands for the spatial curvature and has the values $0,\pm 1$ and $a$ stands for the cosmological scale factor. We have assumed the equation of state $p_m=w\mu_m$ for standard matter, where $w$ is the equation of state parameter.  Note also that the expression connecting
the volume rate of expansion of the fluid $\Theta$ with the Hubble parameter $H$ is given as (see Ref. \cite{amare1})
\begin{equation}
H=\frac{\Theta}{3}.\label{hubble1}
\end{equation}
We proceed with the same approach as presented in Ref.  \cite{scalar5} , where the effective potential $V(\phi)$ is defined in such a way that
\begin{equation}
V'(\phi)=\frac{dV}{d\phi}=\frac{1}{3}\big(2f-(\phi+1)R\big).\label{pot}
\end{equation} 
Now since the background is considered to be the FLRW spacetime, the background  energy density and isotropic pressure quantities for perfect-curvature 
fluid in $f(R)$ theories have the form
\begin{equation}
\mu_{\phi}=\frac{1}{\phi+1}\big[\frac{1}{2}((\phi+1)R-f)-\Theta\dot{\phi}\big],\label{14aa}
\end{equation}
\begin{equation}
p_{\phi}= \frac{1}{\phi+1}\big[\frac{1}{2}(f-R(\phi+1))+\dot{\phi}\dot{\phi}'-\frac{\dot{\phi}^{2}}{\phi'}\phi''+\frac{\phi''\dot{\phi}^{2}}{\phi'^{2}}
  +\frac{2}{3}\Theta \phi'\dot{\phi}\big],\label{14bb}
\end{equation}
and 
\begin{equation}
q^{\phi}_{a}=\pi^{\phi}_{ab}=0.
\end{equation}
The trace of the total EMT is given as
\begin{equation}
T=T^{a}_{a}=3\tilde{p}_{m}-\tilde{\mu}_{m}+T^{\phi}\label{T1aa}
\end{equation}
where the trace of the curvature EMT is  
\begin{equation}
T^{\phi}=\frac{1}{\phi+1}\big[2(f-(\phi+1)R)+3(\Theta \dot{\phi}+\dot{\phi}\dot{\phi}'-\frac{\dot{\phi}^{2}}{\phi'}\phi'' +\phi''\frac{\dot{\phi}^{2}}{\phi'^{2}})\big].\label{T1bb}
\end{equation}
When we consider perfect fluids, the two equations \eqref{fried} and \eqref{duri}  together with the KGE \eqref{KG} are written in the 
scalar field language as
\begin{equation}
\Theta^{2}=\frac{3}{\phi+1}\big[\tilde{\mu}^{m}+\frac{1}{2}((\phi+1)R-f)-\Theta\dot{\phi}\big]-\frac{9K}{a^{2}},\label{fried1}
\end{equation} 
\begin{equation}
\begin{split}
&\dot{\Theta}+\frac{1}{2}\Theta^{2}+\frac{1+3w}{2(\phi+1)}\tilde{\mu}^{m}+\frac{1}{\phi+1}\big[(f-(\phi+1)R)\\
&~~~+\big(3\dot{\phi}'-\frac{3\phi''\dot{\phi}}{\phi'}+\frac{3\phi''\dot{\phi}}{\phi'^{2}}-(2\phi'-1)\Theta\big)\dot{\phi}\big]=0,\label{duri1}
\end{split}
\end{equation}
and 
\begin{equation}
\square \phi-\frac{1}{3}\big[2f-(\phi+1)R+\frac{3w-1}{\phi+1}\tilde{\mu}^{m}\big]=0, \label{KG1}
\end{equation}
respectively.\\
In the following, we will consider some specific toy models of $f(R)$  gravitation.
These models have attracted much attention by cosmologists over the last few decades. We refer the reader to the extensive work done in
Ref. \cite{nojiri2011unified} and more work on inflation analysis for $f(R)$ can be obtained in Ref. \cite{nojiri2003modified,bamba2014inflationary} 
for power-law $f(R)$ model, linear $f(R)$ model and exponential $f(R)$ models. The stability of modified gravity including $f(R)$ 
have been explored extensively in Ref. \cite{carroll410031cosmology}.
 \subsection{The case of $f(R)=\beta R^{n}$ model}

\begin{figure}[pb]
\centerline{\psfig{file=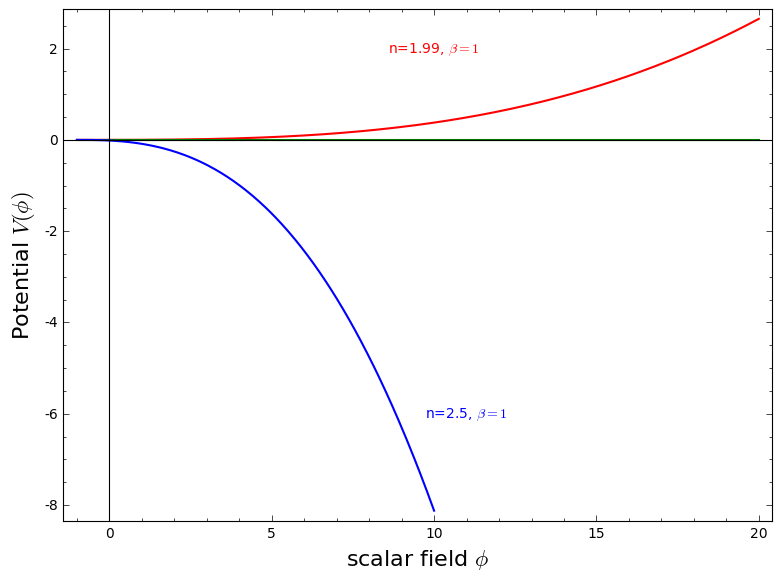,width=10.7cm}}
\vspace*{15pt}
\caption{ The potential $V(\phi)$ in function of the scalar field $\phi$ for $n=1.99,\beta=1$ is 
  colored `red', for $n=2.5,\beta=1$ is colored `blue' .\label{pot11}}
\end{figure}

We consider the model $f(R)=\beta R^{n}$, see Ref. \cite{SanteCarloni3,SanteCarloni2} and we use the relation defined above that $f'=\phi+1$. This model becomes
\begin{equation}
f(\phi)=\beta\big(\frac{\phi+1}{n\beta}\big)^{\frac{n}{n-1}}.\label{fcase1}
\end{equation}
The derivative of the potential,  $V'(\phi)$, is obtained from Eq. \eqref{pot} as  
\begin{equation}
V'(\phi)=\frac{2\beta-n\beta}{3(n\beta)^{n/(n-1)}}(\phi+1)^{n/(n-1)}. \label{v'case1}
\end{equation}
By integrating the above equation with respect to $\phi$, we have the potential $V(\phi)$ 
\begin{equation}
V(\phi)=\frac{\beta(n-1)(2-n)}{3(2n-1)(n\beta)^{n/(n-1)}}(\phi+1)^{(2n-1)/(n-1)}, \label{vcase1}
\end{equation}
where $ n \neq 0,\frac{1}{2}$ and $1$.  The potential $V(\phi)$ dependence on the scalar field $\phi$ 
for the case of the $\beta R^{n}$ toy model is presented from Eq. \eqref{vcase1} in Fig. $\ref{pot11}$ for different values of $n$.
The behavior of the potential for $n=1.99$ is totally different from that of $n=2.5$, when keeping $\beta$ normalized to one. For this toy model,
we have computed the numerical values of spectral index $n_{s}$ and tensor-to-scalar ratio $r$ 
(see definition of there parameters in Section $\ref{sla}$) in the range of the $n\leq 1.99$. We obtained that
the results of tensor-to-scalar ratio $r$ almost fit in the range of Planck data but for spectal index $n_{s}$ tends to unit 
which is not the desired one (see Table $\ref{Tablemodel1}$). 
Then, one can rewrite the Eqs. \eqref{fried1},\eqref{duri1} and \eqref{KG1} as
\begin{equation}
\Theta^{2}=\frac{3}{\phi+1}\big[\tilde{\mu}^{m}-\frac{\beta-n\beta}{2(n\beta)^{n/(n-1)}}(\phi+1)^{n/(n-1)}-\Theta\dot{\phi}\big]-\frac{9K}{a^{2}},\label{friedcase1}
\end{equation}
\begin{equation}
\begin{split}
&\dot{\Theta}+\frac{1}{3}\Theta^{2}+\frac{1+3w}{2(\phi+1)}\tilde{\mu}^{m}+\Big\{\frac{\beta-n\beta}{(n\beta)^{\frac{n}{(n-1)}}}(\phi+1)^{\frac{n}{(n-1)}}\\
&+\frac{\dot{\phi}}{\phi+1}\Big(3(n-2)\dot{\phi}[\frac{(\phi+1)^{-1}}{n+1}-\big(\frac{\phi+1}{n\beta}\big)^{\frac{-1}{(n-1)}}]-\Theta\big)\\
&+(n-1)(\phi+1)^{\frac{(n-2)}{(n-1)}}[3(n\beta)^{\frac{1}{(n-1)}}\ddot{\phi}+2(n\beta)^{\frac{-1}{(n-1)}}\Theta]\Big)\Big\}=0,\label{duricase1}
\end{split}
\end{equation}
and 
\begin{equation}
\square \phi-\frac{1}{3}\big[\frac{2\beta-n\beta}{(n\beta)^{n/(n-1)}}(\phi+1)^{n/(n-1)}+\frac{3w-1}{\phi+1}\tilde{\mu}^{m}\big]=0,\label{KGcase1}
\end{equation}
respectively. Eq. \eqref{friedcase1} is quadratic in the volume rate of expansion of the fluid $\Theta$. Thus the roots $\Theta_{1}$ and $\Theta_{2}$ are given as
\begin{equation}
\Theta_{1,2}=-\frac{3\dot{\phi}}{2(\phi+1)}\pm \frac{1}{2}\sqrt{\Big(\frac{3\dot{\phi}}{(\phi+1)}\Big)^{2}-\Big[\frac{12}{(\phi+1)}Z+\frac{36K}{a^{2}}\Big]},\label{theta1case1}
\end{equation}
where $Z=\big(\frac{\beta(1-n)}{2(n\beta)^{n/(n-1)}}(\phi+1)^{n/(n-1)}-\tilde{\mu}^{m}\big)$.

\begin{figure}[pb]
\centerline{\psfig{file=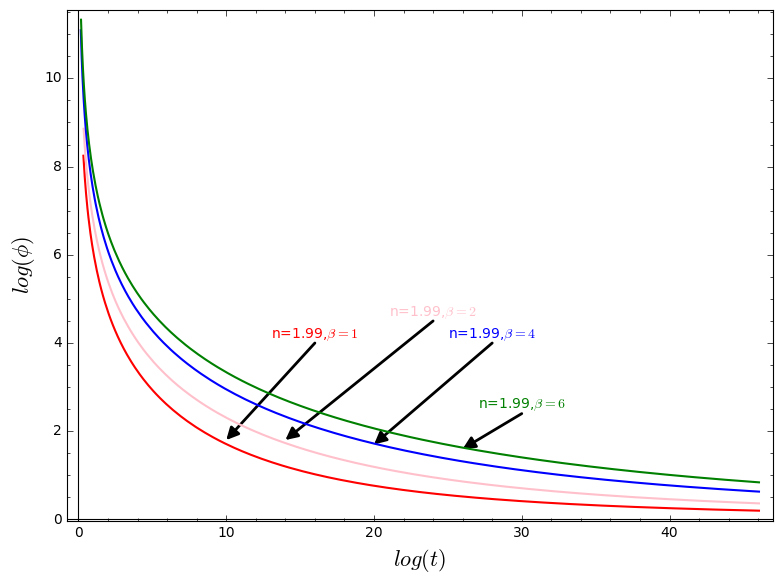,width=10.7cm}}
\vspace*{15pt}
\caption{ The scalar field $\phi(t)$ in function of time $t$ for $n=1.99,\beta=1$ is colored `red', for $n=1.99,\beta=2$ is 
  colored `pink', for $n=1.99,\beta=4$ is colored `blue' and for $n=1.99,\beta=6$ is colored `green'.\label{pix1}}
\end{figure}

The Hubble parameter defined in Eq. \eqref{hubble1} is presented according to the above roots as
\begin{equation}
H_{1,2}=-\frac{\dot{\phi}}{2(\phi+1)}\pm \frac{1}{6}\sqrt{\Big(\frac{3\dot{\phi}}{(\phi+1)}\Big)^{2}-\Big[\frac{12}{(\phi+1)}Z+\frac{36K}{a^{2}}\Big]}.\label{hubble1case1}
\end{equation}
When we consider Eq. \eqref{14aa} and use Eq. \eqref{fcase1}, we have the energy density $\mu_{\phi}$ given as
\begin{equation}
\mu_{\phi}=\frac{\beta}{2}\frac{(n-1)}{(n\beta)^{n/(n-1)}}(\phi+1)^{1/(n-1)}-\Theta\frac{\dot{\phi}}{\phi+1}.\label{dencase1}
\end{equation}
As  has been mentioned in the discussion about the BD theory, it is clear that the scalar field couples with the matter through the matter EMT.
This expression of scalar energy density in Equations \eqref{dencase1} together with $\tilde{\mu}_{m}=\frac{\mu_{m}}{\phi+1}$ shows the direct implications 
of the scalar field in the matter from the total energy density $\mu=\mu_{\phi}+\tilde{\mu}_{m}$. From the expression of the perfect fluid, we have also the
dependence of total pressure $p$ to the scalar field $\phi$ through $p=w \mu$, where $w$ is the equation of state parameter as stated in the previous sections.
Considering Eq. \eqref{KGcase1}, since the scalar field $\phi$ is assumed to be time dependent only (as stressed in Ref. \cite{dicke2}), 
we drop out the spatial dependence on the covariant d'Alembert operator and only the second-order time derivative will remain. Thus we have
\begin{equation}
-\ddot{\phi}-\frac{1}{3}\Big[\frac{\beta(2-n)}{(n\beta)^{n/(n-1)}}(\phi+1)^{n/(n-1)}+\frac{3w -1}{\phi+1}\tilde{\mu}_{m}\Big]=0.\label{kkg1}
\end{equation} 
In the early universe matter  energy density can be neglected over the scalar field.  One can therefore neglect the $\tilde{\mu}_{m}$ term in Eq. \eqref{kkg1} and rewrite
\begin{equation}
\ddot{\phi}+\frac{1}{3}\frac{\beta(2-n)}{(n\beta)^{n/(n-1)}}(\phi+1)^{n/(n-1)}=0.
\end{equation}
This is a nonlinear differential equation in $\phi$. This equation is an autonomous second-order differential equation.
Let us define $b=\frac{1}{3}\frac{\beta(2-n)}{(n\beta)^{n/(n-1)}}$ and $\lambda=n/(n-1)$, the equation takes the form
\begin{equation}
\ddot{\phi}+b(\phi+1)^{\lambda}=0.
\end{equation} 
We define $y$ to be
\begin{equation}
y=\frac{d\phi}{dt}, \text{ and } \frac{dy}{dt}=\frac{dy}{d\phi}\frac{d\phi}{dt}=\frac{dy}{d\phi}y.
\end{equation}
Thus we write our differential equation as
\begin{equation}
y\frac{dy}{d\phi}+b(\phi+1)^{\lambda}=0. \label{shortdiff}
\end{equation}
This differential equation has the solution 
\begin{equation}
\phi(t)=\Big(\frac{2\sqrt{2b/(\lambda-1)}}{1-\lambda}\Big)^{2/(1-\lambda)} t^\frac{2}{(1-\lambda)}-1.
\end{equation}
Replacing back the expressions for $\lambda$ and $b$, we have
\begin{equation}
\phi(t)=\Bigg[2(1-n)\sqrt{\frac{2\beta(2-n)(1-n)}{3(2n-1)(n\beta)^{n/(n-1)}}}\Bigg]^{2(1-n)}t^{2(1-n)}-1. \label{sol1}
\end{equation} 
At this point, one can have an interest in obtaining the dependence on time of the energy density $\mu_{\phi}$ presented in Eq. \eqref{dencase1}.
The same interest may be extended to the volume expansion rate presented in Eq. \eqref{theta1case1}, but this will require to specify the 
curvature $K$ and the scale factor $a$.
Here we present the solution of the KGE obtained in Eq. \eqref{sol1} in Fig. $\ref{pix1}$. Keeping $n$ constant and allowing the 
value of $\beta$ to vary, we observe the change. As $\beta$ increases, the slope is becoming less steep. This clearily shows that the scalar field 
was dominating at early stage and less dominant in the late stage.

 \subsection{The case of $f(R)=\alpha R+\beta R^{n}$ model }

Among the most widely studied $f(R)$ models are the  $f(R)=\alpha R+\beta R^{n}$ (see Ref. \cite{SanteCarloni2}). These models can be rewritten as
\begin{equation}
f(\phi)=\alpha\big(\frac{\phi+1-\alpha}{n\beta}\big)^{1/(n-1)}+\beta\big(\frac{\phi+1-\alpha}{n\beta}\big)^{n/(n-1)}.\label{fcase2}
\end{equation}
The expression for the derivative of the  potential, $V'(\phi)$, has the form
\begin{equation}
\begin{split}
V'(\phi)=&\frac{1}{3}\big[(2\alpha-(\phi+1))\big(\frac{\phi+1-\alpha}{n\beta}\big)^{1/(n-1)}\\
&+2\beta \big(\frac{\phi+1-\alpha}{n\beta}\big)^{n/(n-1)}\big].\label{v'case2}
\end{split}
\end{equation}
Thus the potential can be given by
\begin{equation}
\begin{split}
V(\phi)=&\frac{(1-n)(\alpha-\phi-1)\big[n(3\alpha-\phi-1)-3\alpha+2(\phi+1)\big]}{3n(2n-1)}\times\\
&\frac{\big(\frac{\phi+1-\alpha}{n\beta}\big)^{1/(n-1)}}{3n(2n-1)}, \text{ n $\neq$ 0,$\frac{1}{2}$ and 1}.\label{vcase2}
\end{split}
\end{equation}

\begin{figure}[pb]
\centerline{\psfig{file=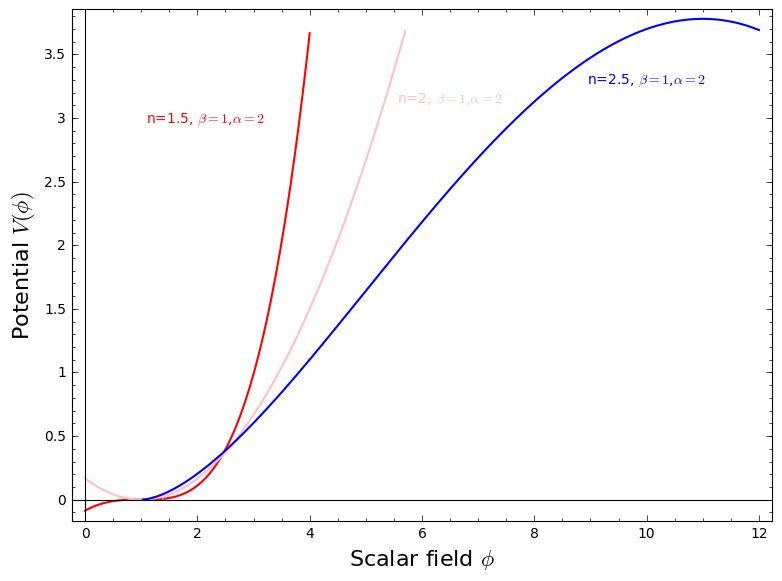,width=10.7cm}}
\vspace*{15pt}
\caption{ For the model of $f(R)=\alpha R+\beta R^{n}$, the potential $V(\phi)$ in function of the scalar field $\phi$ for
$n=1.5,\beta=1,\alpha=2$ is colored `red', for $n=2,\beta=2,\alpha=2$ is colored `pink' and for $n=2.5,\beta=3,\alpha=2$ is
colored `blue' .\label{pot21}}
\end{figure}

The potential $V(\phi)$ dependence on the scalar field $\phi$ for the case of the $\alpha R+\beta R^{n}$ is presented from 
Eq. \eqref{vcase2} in Fig. $\ref{pot21}$.
The behavior of the potential for $n=1.5$, $n=2$ are parabola with upwards concavity. For $n=2.5$, the concavity has been changed
and the maximum is located at a finite point. Thus the transition point is located in the interval $2<n<2.1$. The inflationary stability
only supports the shape  of the potential to be in $U$ form. Therefore the potential in `red' and the one in `pink'  obey the stability requirement. 
This constrains the power $n$ (which is the power of the $f(R)$ model) to be less than $2.5$. The numerical computations of $n_{s}$ and $r$ for just
$n=1.5$ have shown to fit in the Planck date (see Table $\ref{Tablemodel2}$).
 
\begin{figure}[pb]
\centerline{\psfig{file=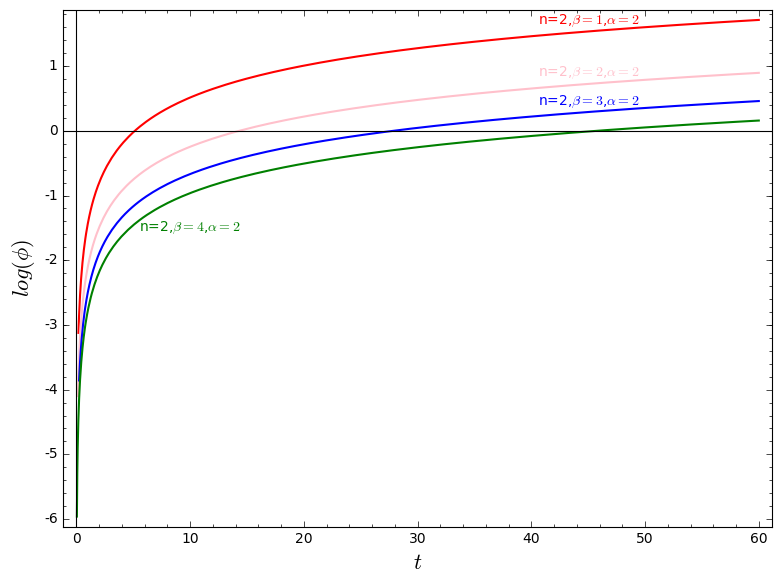,width=10.7cm}}
\vspace*{15pt}
\caption{ The scalar field $\phi(t)$ in function of time $t$ for $n=1.99,\beta=1$ is colored `red', for $n=1.99,\beta=2$ is colored
`pink', for $n=1.99,\beta=4$ is colored `blue' and for $n=1.99,\beta=6$ is colored `green'.\label{pix2}}
\end{figure}

In this model, we write the Eqs. \eqref{fried1},(\ref{duri1} and \eqref{KG1} as
\begin{equation}
\Theta^{2}=\frac{3}{\phi+1}\big[\tilde{\mu}^{m}-\frac{\beta-n\beta}{2(n\beta)^{n/(n-1)}}(\phi+1-\alpha)^{n/(n-1)}-\Theta\dot{\phi}\big]-\frac{9K}{a^{2}},\label{friedcase2}
\end{equation}
\begin{equation}
\begin{split}
&\dot{\Theta}+\frac{1}{3}\Theta^{2}+\frac{1+3w}{2(\phi+1)}\tilde{\mu}^{m}+\frac{\beta-n\beta}{(n\beta)^{\frac{n}{(n-1)}}}(\phi+1-\alpha)^{\frac{n}{(n-1)}}\\
&+\frac{\dot{\phi}}{\phi+1}\Big(3(n-2)\dot{\phi}[\frac{(\phi+1-\alpha)^{-1}}{n+1}-\big(\frac{\phi+1-\alpha}{n\beta}\big)^{\frac{-1}{(n-1)}}]-\Theta\big)\\
&+(n-1)(\phi+1-\alpha)^{\frac{(n-2)}{(n-1)}}[3(n\beta)^{\frac{1}{(n-1)}}\ddot{\phi}+2(n\beta)^{\frac{-1}{(n-1)}}\Theta]\Big)=0,
\label{duricase2}
\end{split}
\end{equation}
and
\begin{equation}
\begin{split}
&\square \phi -\frac{1}{3}\big[-\frac{(\phi+1-2\alpha)}{(n\beta)^{1/(n-1)}}(\phi+1-\alpha)^{1/(n-1)}\\
&+\frac{2\beta}{(n\beta)^{n/(n-1)}}(\phi+1-\alpha)^{n/(n-1)}+\frac{3w-1}{\phi+1}\tilde{\mu}^{m}\big]=0,\label{KGcase2}
\end{split}
\end{equation}
respectively.
 Thus the roots of Eq. \eqref{friedcase2}, $\Theta_{1}$ and $\Theta_{2}$, are given as
\begin{equation}
\Theta_{1,2}=-\frac{3\dot{\phi}}{2(\phi+1)}\pm \frac{1}{2}\sqrt{\Big(\frac{3\dot{\phi}}{(\phi+1)}\Big)^{2}-\Big[\delta+\frac{36K}{a^{2}}\Big]},\label{theta1case2}
\end{equation}
where $\delta = \frac{12}{(\phi+1)}\big(\frac{\beta(1-n)}{2(n\beta)^{n/(n-1)}}(\phi+1-\alpha)^{n/(n-1)}-\tilde{\mu}^{m}$.
The Hubble parameter defined in Eq. \eqref{hubble1} is given according to the above roots as
\begin{equation}
H_{1,2}=-\frac{\dot{\phi}}{2(\phi+1)}\pm \frac{1}{6}\sqrt{\Big(\frac{3\dot{\phi}}{(\phi+1)}\Big)^{2}-\Big[\delta+\frac{36K}{a^{2}}\Big]}.\label{hubble1case2}
\end{equation} 
Using Eq. \eqref{fcase2} in Eq. \eqref{14aa}, we have the energy density $\mu_{\phi}$ as
\begin{equation}
\mu_{\phi}=\frac{1}{\phi+1}\Big[\frac{(n-1)}{2n(n\beta)^{n/(n-1)}}\big(\phi+1-\alpha\big)^{(2n-1)/(n-1)}-\Theta\dot{\phi}\Big].\label{dencase2}
\end{equation}
Dropping out the spatial dependence in the covariant
d'Alembert operator in Eq. \eqref{KGcase2} yields
\begin{equation}
\begin{split}
&-\ddot{\phi}-\frac{1}{3}\Big[-\frac{(\phi+1-2\alpha)}{(n\beta)^{1/(n-1)}}(\phi+1-\alpha)^{1/{n-1}}\\
&+\frac{2\beta}{(n\beta)^{n/(n-1)}}(\phi+1-\alpha)^{n/(n-1)}+\frac{3w-1}{\phi+1}\tilde{\mu}^{m}\Big]=0.
\end{split}
\end{equation}
As stated in the previous section, we neglect the energy density $\tilde{\mu}_{m}$ to obtain
\begin{equation}
\begin{split}
&\ddot{\phi}+\frac{1}{3}\Big[\frac{(\phi+1-2\alpha)}{(n\beta)^{1/(n-1)}}(\phi+1-\alpha)^{1/{n-1}}\\
&-\frac{2\beta}{(n\beta)^{n/(n-1)}}(\phi+1-\alpha)^{n/(n-1)}\Big]=0.\label{kkg2}
\end{split}
\end{equation}
If we use the same approach used for the autonomous equation presented in the previous section (see Eq. \eqref{kkg1}), we obtain the solution to this equation after using the Taylor approximation truncated at the order of $O(\phi^{3})$ during the integration process and it was assumed that $\phi$ is small enough so that the $f(R)$ under consideration is as close to GR as possible. The solution obtained is the scalar field dependence on time and it is presented below:
\begin{equation}
\phi=\frac{2m_{0}^{3/2}}{m_1}\Big[-\frac{1}{\sqrt{m_{0}}}\pm \sqrt{\frac{1}{m_{0}}+\frac{m_{1}}{\sqrt{m_{2}}m_{0}^{3/2}}\Big(\ln (2\sqrt{m_{0}m_{2}}+m_{1})\Big)t}\Big]\;,
\end{equation}
where constants $m_{0},m_{1}$ and $m_{2}$ are defined as follows:
\begin{equation*}
\begin{split}
m_{0}&=\frac{2\beta(n-1)(1-\alpha)^{(2n-1)/(n-1)}}{3(2n-1)(n\beta)^{n/(n-1)}}\\
&-\frac{(1-\alpha)^{(2-n)/(n-1)}\big(-2\alpha+1-(3\alpha-1)(n-1)\big)}{3n(2n-1)(n\beta)^{1/(n-1)}},
\end{split}
\end{equation*}
\begin{equation*}
m_{1}=(1-\alpha)^{n/(n-1)}+\frac{(2\alpha-1)(1-\alpha)^{1/(n-1)}}{3(n\beta)^{1/(n-1)}}
\end{equation*}
and 
\begin{equation*}
\begin{split}
m_{2}=\frac{\beta(1-\alpha)^{1/(n-1)}}{3(n\beta)^{n/(n-1)}}
+\frac{\big((2\alpha-1)+(\alpha+1)(n-1)\big)(1-\alpha)^{(2-n)/(n-1)}}{6(n-1)(n\beta)^{1/(n-1)}}.
\end{split}
\end{equation*}
We have used the motivation that as $\phi=f'-1$, for the $f(R)$ under consideration to be reduced to  GR, $\phi$ has to be infinitesimally small. As mentioned in the previous section, it is possible to obtain the dependence on time of the energy density $\mu_{\phi}(t)$ presented in Eq. \eqref{dencase2}. We can also obtain the time dependence of the volume expansion rate $\Theta$ presented in Eq. \eqref{theta1case2}, as stressed in the previous section, the specification of the curvature $K$ and the scale factor $a$ is required to go through. 
Here we present the scalar field $\phi$ dependence on time $t$ in Fig. $\ref{pix2}$. It can easily be observed that as we stress $n$ to be constant and keep $\beta$ changing while still keeping $\alpha$ constant, the function $\phi(t)$ crosses $t$-axis at different values. Note that in that context, the slope is nearly the same for all curves but the position is shifted downwards as $\beta$ increases.

 \subsection{The case of $f(R)=R-\frac{\nu^{4}}{R}$ model }

\begin{figure}[pb]
\centerline{\psfig{file=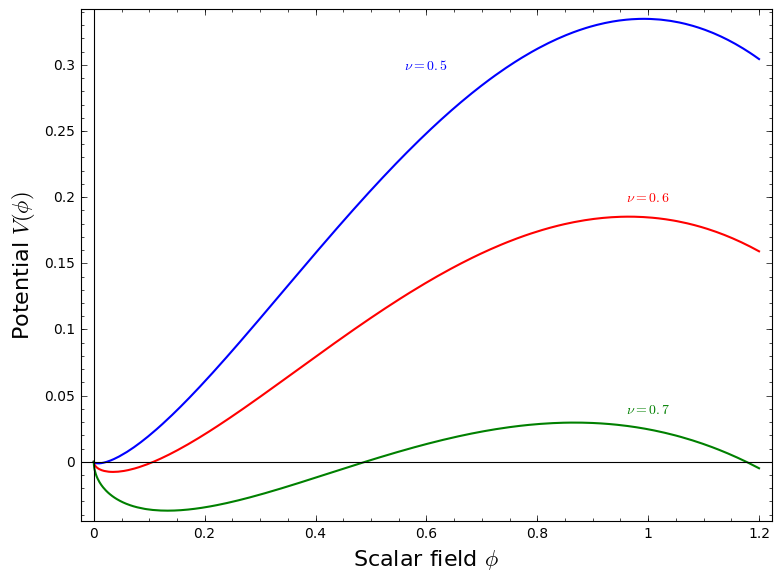,width=10.7cm}}
\vspace*{15pt}
\caption{ For the model of $f(R)=R-\frac{\nu^{4}}{R}$, the potential $V(\phi)$ in function of the scalar field $\phi$ for $\nu=0.5$
  is colored `blue', for $\nu=0.6$ is colored `red' and for $\nu=0.7$ is colored `green'.\label{pot3a}}
\end{figure}

\begin{figure}[pb]
\centerline{\psfig{file=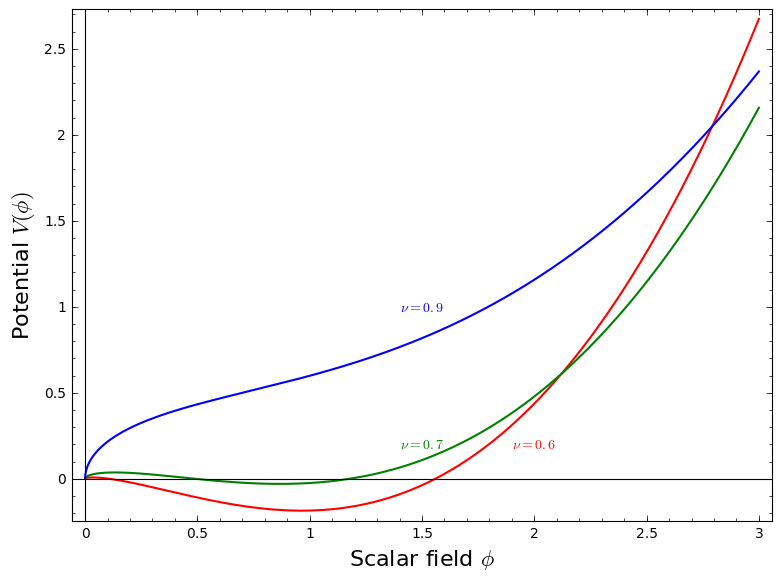,width=10.7cm}}
\vspace*{15pt}
\caption{ For the model of $f(R)=R-\frac{\nu^{4}}{R}$, with $R_{-}$, the potential $V(\phi)$ in function of the scalar field $\phi$ 
  for $\nu=0.9$ is colored `blue', for $\nu=0.6$ is colored `red' and for $\nu=0.7$ is colored `green'.\label{pot3b}}
\end{figure}

This model has been studied by several authors (see, for example, Ref. \cite{scalar3,Carroll1,carroll410031cosmology}). The model reduces to GR when $\nu=0$. 
In this model, the Ricci scalar as function of $\phi$ is given as
\begin{equation}
R(\phi)=\pm \frac{\sqrt{\phi}}{\nu^{2}}. \label{Rcase3}
\end{equation}
And the function $f(\phi)$ will depend on the sign of $R(\phi)$ in Eq. \eqref{Rcase3}. When considering positive sign, we write the function $f(\phi)$ as
\begin{equation}
f(\phi)=(\frac{1}{\nu^{2}}-\frac{\nu^{6}}{\phi})\sqrt{\phi}.\label{fcase3a}
\end{equation}
Thus, the corresponding potential $V'(\phi)$ is given as
\begin{equation}
V'(\phi)=\frac{1}{3}\Big(\frac{(1-\phi)}{\nu^{2}}-\frac{2\nu^{6}}{\phi}\Big)\sqrt{\phi}. \label{v'case3a}
\end{equation}
The potential is obtained by integrating Eq. $\ref{v'case3a}$ to get 
\begin{equation}
V(\phi)=\frac{1}{3}\Big(\frac{2\phi^{3/2}}{3\nu^{2}}-\frac{2\phi^{5/2}}{5\nu^{2}}-4\nu^{6}\phi^{1/2}\Big) .\label{vcase3a}
\end{equation}
The potential $V(\phi)$ dependence on the scalar field $\phi$ for the case of the $R-\frac{\nu^{4}}{R}$
is presented  in Fig. $\ref{pot3a}$ for the Ricci scalar with positive root. One can easily see that as the value of $\nu$ is getting bigger, the 
turning point approaches the horizontal axis. This implies that the value of $\nu$ influences the stability behavior of 
the potential $V(\phi)$. For this model to have a $U$ form (which is the preferred form for the inflation potential), the value of $\nu$
has to be constrained to be less and less. By considering the parameter $\nu$ to be $0.5$ and $0.6$, we have computed numerically the corresponding
values of $n_{s}$ and $r$ and have shown to be almost fit with Planck data (see Table $\ref{Tablemodel3a}$).
  In this model, we write the KGE \eqref{KG1} as
\begin{equation}
\square \phi -\frac{1}{3}\Big[(\frac{1}{\nu^{2}}-\frac{\phi}{\nu^{2}}-2\nu^{6}\phi^{-1})\sqrt{\phi}+\frac{3w-1}{\phi+1}\tilde{\mu}^{m}
\Big]=0\;.\label{KGcase3a}
\end{equation}
By dropping out the spatial dependence in the covariant 
d' Alembert operator in Eq. \eqref{KGcase3a} the KGE becomes
\begin{equation}
-\ddot{\phi}-\frac{1}{3}\Big[(\frac{1}{\nu^{2}}-\frac{\phi}{\nu^{2}}-2\nu^{6}\phi^{-1})\sqrt{\phi}+\frac{3w-1}{\phi+1}\tilde{\mu}^{m}
\Big]=0.
\end{equation}
Once again assuming that in the early universe 
matter had much smaller effect compared to the scalar field, we reduce the KGE to
\begin{equation}
\ddot{\phi}+\frac{1}{3}\Big[(\frac{1}{\nu^{2}}-\frac{\phi}{\mu^{2}}-2\nu^{6}\phi^{-1})\sqrt{\phi}\Big]=0\;,\label{kkg3a}
\end{equation}
the solution to which can be given by
\begin{equation}
\frac{4\phi^{3/4}}{3\sqrt{-b}}-\frac{2a\phi^{7/4}}{7(-b)^{3/2}}+\frac{(3a^{2}-4bc)\phi^{11/4}}{22(-b)^{5/2}}+O(\phi^{15/4})=t, 
\end{equation}
where $a=\frac{4}{9\nu^{2}}$, $b=\frac{8\nu^{6}}{3}$ and $c=\frac{4}{15\nu^{2}}$.
When considering negative sign in Equation representing Ricci scalar dependence on scalar field $\phi$, we write the function $f(\phi)$ as
\begin{equation}
f(\phi)=(-\frac{1}{\nu^{2}}+\frac{\nu^{6}}{\phi})\sqrt{\phi}.\label{fcase3b}
\end{equation}
Thus, the corresponding potential $V'(\phi)$ is given as
\begin{equation}
V'(\phi)=\frac{1}{3}\Big(-\frac{\phi^{1/2}}{\nu^{2}}+2\nu^{6}\phi^{-1/2}+\frac{\phi^{3/2}}{\nu^{2}}\Big). \label{v'case3b}
\end{equation}
The potential is obtained by integrating Eq. $\ref{v'case3b}$ to get 
\begin{equation}
V(\phi)=\frac{1}{3}\Big(-\frac{2\phi^{3/2}}{3\nu^{2}}+\frac{2\phi^{5/2}}{5\nu^{2}}+4\nu^{6}\phi^{1/2}\Big) .\label{vcase3b}
\end{equation}
The potential $V(\phi)$ dependence on the scalar field $\phi$ for  $f(R)=R-\frac{\nu^{4}}{R}$, with negative sign on the Ricci scalar root, 
is presented from Eq. \eqref{vcase3a} in Fig. $\ref{pot3b}$. In this figure, as one can see, when the value of $\nu$ is getting
bigger the potential is becoming wider. When the value of $\nu$ is getting less, the potential is shaping into a $U$ form.
Thus one might see that like in the case of the positive root (see Fig. $\ref{pot3a}$), the same inspection can be drawn that as the 
value of $\nu$ is small the potential $V(\phi)$ is becoming the one which supports inflation. By considering the parameter $\nu$ to be $0.9$,
we have computed numerically the corresponding values of $n_{s}$ and $r$ and have shown to fit with Planck data (see Table $\ref{Tablemodel3b}$).
 
  In this model, the KGE \eqref{KG1} reduces to
\begin{equation}
\square \phi -\frac{1}{3}\Big[-\frac{\phi^{1/2}}{\nu^{2}}+2\nu^{6}\phi^{-1/2}+\frac{\phi^{3/2}}{\nu^{2}}+\frac{3w-1}{\phi+1}\tilde{\mu}^{m}
\Big]=0\;,\label{KGcase3b}
\end{equation}
which simplifies further to
\begin{equation}
-\ddot{\phi} -\frac{1}{3}\Big[-\frac{\phi^{1/2}}{\nu^{2}}+2\nu^{6}\phi^{-1/2}+\frac{\phi^{3/2}}{\nu^{2}}+\frac{3w-1}{\phi+1}\tilde{\mu}^{m}
\Big]=0\label{KGcase3c}
\end{equation}
when spatial variations are neglected. Following similar arguments as before, we decouple the scalar field from matter and rewrite 
\begin{equation}
\ddot{\phi}+\frac{1}{3}\Big[-\frac{\phi^{1/2}}{\mu^{2}}+2\nu^{6}\phi^{-1/2}+\frac{\phi^{3/2}}{\nu^{2}}\Big]=0\;.\label{kkg3b}
\end{equation}
A truncated solution to this equation can be given by
\begin{equation}
\frac{4\phi^{3/4}}{3(b)^{1/2}}+\frac{2a\phi^{7/4}}{7(b)^{3/2}}+\frac{(3a^{2}-4bc)\phi^{11/4}}{22(b)^{5/2}}+O(\phi^{15/4})=t\;,
\end{equation}
where $a=\frac{4}{9\nu^{2}}$, $b=\frac{8\nu^{6}}{3}$ and $c=\frac{4}{15\nu^{2}}$.

 \subsection{The case of $f(R)=R-(1-m)\nu^{2}\Big(\frac{R}{\nu^{2}}\Big)^{m}-2\Lambda$ model }

\begin{figure}[pb]
\centerline{\psfig{file=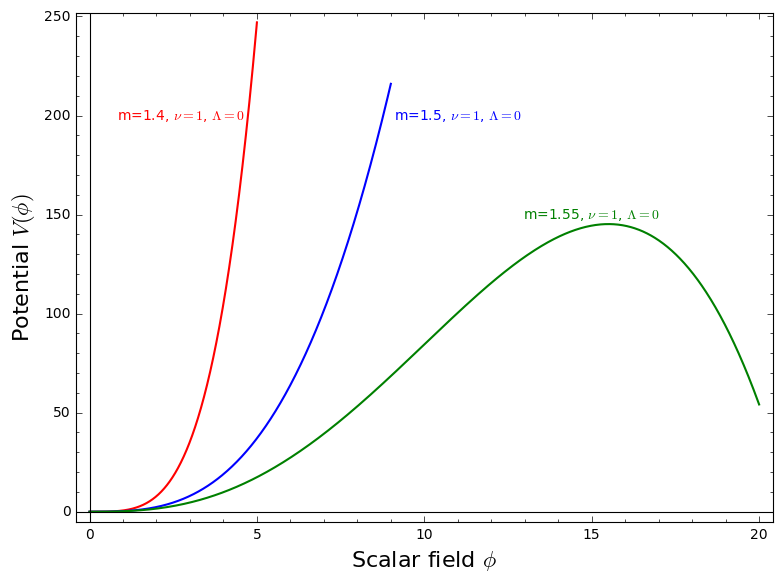,width=10.7cm}}
\vspace*{15pt}
\caption{ For the model of $f(R)=R-(1-m)\nu^{2}\Big(\frac{R}{\nu^{2}}\Big)^{m}-2\Lambda$, the potential $V(\phi)$ in function of
  the scalar field $\phi$ for $m=1.4,\nu=1, \Lambda=0$ is colored `red', for $m=1.5,\nu=1, \Lambda=0$ is colored `blue' and for 
  $m=1.55,\nu=1, \Lambda=0$ is colored `green'.\label{pot4}}
\end{figure}

This model can be considered as the generalization of the $f(R)=R-\frac{\nu^{4}}{R}$ discussed above, see Ref. \cite{scalar3,Amendola1} for more details
and extensive work about this model type.
With this model, the Ricci scalar as a function of $\phi$ is given as
\begin{equation}
R(\phi)=\frac{\nu^{2}\phi^{1/(m-1)}}{[m(m-1)]^{1/(m-1)}}. \label{Rcase4}
\end{equation}
We use the expression of $R(\phi)$ in Eq. \eqref{Rcase4} to obtain the function $f(\phi)$  as 
\begin{equation}
f(\phi)=\frac{\nu^{2}\phi^{1/(m-1)}}{[m(m-1)]^{1/(m-1)}}-\frac{(1-m)\nu^{2}\phi^{m/(m-1)}}{[m(m-1)]^{m/(m-1)}}-2\Lambda.\label{fcase4}
\end{equation}
Thus
\begin{equation}
\begin{split}
V'(\phi)=&\frac{1}{3}\Big[\frac{\nu^{2}\phi^{1/(m-1)}}{[m(m-1)]^{1/(m-1)}}-\frac{2(1-m)\nu^{2}\phi^{m/(m-1)}}{[m(m-1)]^{m/(m-1)}}\\
&-\frac{\nu^{2}\phi^{m/(m-1)}}{[m(m-1)]^{1/(m-1)}}-4\Lambda \Big], \label{v'case4}
\end{split}
\end{equation}
and therefore
\begin{equation}
\begin{split}
V(\phi)=&\frac{1}{3}\Big[\frac{(m-1)\nu^{2}\phi^{m/(m-1)}}{m[m(m-1)]^{1/(m-1)}}+
\frac{2(m-1)^{2}\nu^{2}\phi^{(2m-1)/(m-1)}}{(2m-1)[m(m-1)]^{m/(m-1)}}\\
&-4\Lambda\phi-\frac{(m-1)\nu^{2}\phi^{(2m-1)/(m-1)}}{(2m-1)[m(m-1)]^{1/(m-1)}}\Big] .\label{vcase4}
\end{split}
\end{equation}
The potential $V(\phi)$ dependence on the scalar field $\phi$ for the case of the $R-(1-m)\nu^{2}\Big(\frac{R}{\nu^{2}}\Big)^{m}-2\Lambda$ 
is depicted in Fig. $\ref{pot4}$. We have set the value of $\lambda=0$ and normalized the parameter $\nu$ to unity. 
Thus we allowed the parameter $m$ to change so that we can easily observe how it affects the model. 
From this figure, one can see that as the value of $m$ is becoming bigger, the shape is also become wider. But this  stops before reaching
$m=1.55$. There is a transition between $1.5<m<1.55$. Therefore the potential in $U$ from is only supported for the potential $V(\phi)$
with $m<1.55$. Thus the shape of the inflation potential  puts constraint on this $f(R)$ model. We have normalized $\nu=1$ and computed numerical
values of $n_{s}$ and $r$ for different values of $m$. One can see from Table $\ref{Tablemodel4}$ that the obtained values fit with Planck data.
  In this model, we write KGE as
\begin{equation}
\begin{split}
\square \phi -&\frac{1}{3}\Big[\frac{\nu^{2}\phi^{1/(m-1)}}{[m(m-1)]^{1/(m-1)}}-\frac{2(1-m)\nu^{2}\phi^{m/(m-1)}}{[m(m-1)]^{m/(m-1)}}\\
&-\frac{\nu^{2}\phi^{m/(m-1)}}{[m(m-1)]^{1/(m-1)}}-4\Lambda +\frac{3w-1}{\phi+1}\tilde{\mu}^{m}
\Big]=0\;,\label{KGcase4}
\end{split}
\end{equation}
and following similar arguments of neglecting spatial derivatives as in the previous models, we have
\begin{equation}
\begin{split}
-\ddot{\phi}-&\frac{1}{3}\Big[\frac{\nu^{2}\phi^{1/(m-1)}}{[m(m-1)]^{1/(m-1)}}-\frac{2(1-m)\nu^{2}\phi^{m/(m-1)}}{[m(m-1)]^{m/(m-1)}}\\
-&\frac{\nu^{2}\phi^{m/(m-1)}}{[m(m-1)]^{1/(m-1)}}-4\Lambda+\frac{3w-1}{\phi+1}\tilde{\mu}^{m}
\Big]=0.
\end{split}
\end{equation}
Decoupling the scalar field from matter, one gets \begin{equation}
\begin{split}
\ddot{\phi}+&\frac{1}{3}\Big[\frac{\nu^{2}\phi^{1/(m-1)}}{[m(m-1)]^{1/(m-1)}}-\frac{2(1-m)\nu^{2}\phi^{m/(m-1)}}{[m(m-1)]^{m/(m-1)}}\\
&-\frac{\nu^{2}\phi^{m/(m-1)}}{[m(m-1)]^{1/(m-1)}}-4\Lambda\Big]=0\;,\label{kkg4}
\end{split}
\end{equation}
with the solution \begin{equation}
-\frac{(2-2m)\phi\Big(\frac{A+B\phi}{B\phi}\Big)^{1/2}F_{1}(\frac{1}{2},\frac{1}{2(m-1)};\frac{2m-1}{2(m-1)};-\frac{A}{B\phi})}{
\Big(\phi^{m/(m-1)}(A+B\phi)\Big)^{1/2}}=t, 
\end{equation}
where $A=\frac{2(1-m)\nu^{2}}{3m[m(m-1)]^{1/(m-1)}}$,\\
$B=\frac{2(m-1)}{3(2m-1)}\Big(\frac{\nu^{2}}{[m(m-1)]^{1/(m-1)}}-
\frac{2(m-1)\nu^{2}}{[m(m-1)]^{m/(m-1)}}\Big)$ and $F_{1}(k,l;s;x)$ is the hypergeometric function.\\
Setting $m=\frac{3}{2}$ as we did when $V(\phi)$ was plotted in Fig. $\ref{pot4}$, we have
\begin{equation}
\frac{\sqrt{81}}{\sqrt{-8}\phi^{5/2}}\sum^{\infty}_{n\geq 0}\frac{\Gamma(\frac{1}{2}+n)}{\sqrt{\pi}}\frac{1}{n!}\left(\frac{4}{\phi}\right)^{n} =t,
\end{equation}
where $\Gamma(z)$ is the gamma function.

  \section{Slow-roll Approximation}\label{sla}

The slow-roll approach to inflation  has attracted many cosmologists in studying inflation dynamics (see Ref. \cite{slow3,slow1,slow2}).
There are two conditions leading to slow-rolling \cite{slow1}. The first one says that the square of the time derivative of the slow-rolling scalar
field has to be smaller than the slow-rolling scalar field potential. This is mathematically
\begin{equation}
\big(\frac{d\phi}{dt}\big)^{2}<V(\phi). \label{slow1}
\end{equation}
The second condition is about the second-order time derivative which is conditioned to be smaller than the derivative of the potential with 
respect to the scalar field $\phi$. This is
\begin{equation}
2\big|\frac{d^{2}\phi}{dt^{2}}\big|<|V'(\phi)|.\label{slow2}
\end{equation}
At this moment, we define two parameters $\epsilon (\phi)$ and $\eta(\phi)$: they are called potential slow-roll parameters. Their expressions are (see Ref. \cite{slow2}):
\begin{equation}
\epsilon(\phi)=\frac{1}{2\kappa^{2}}\Big(\frac{V'(\phi)}{V(\phi)}\Big)^{2},
\end{equation}
and
\begin{equation}
\eta(\phi)=\frac{1}{\kappa^{2}}\Big(\frac{V''(\phi)}{V(\phi)}\Big)\;.
\end{equation}
The spectral index $n_{s}$ and the tensor-to-scalar ratio $r$ are defined in Ref. \cite{slow3,slow4} as 
\begin{equation}
n_{s}=1-6\epsilon+2\eta,\label{ns}
\end{equation}
and 
\begin{equation}
r=16\epsilon\label{r}
\end{equation}
respectively.

 \subsection{Slow-roll approximation of $f(R)=\beta R^{n}$}

For the case of $f(R)=\beta R^{n}$, the parameters $\epsilon(\phi)$ and $\eta(\phi)$ take the form
\begin{equation}
\epsilon(\phi)=\frac{(2n-1)^{2}}{2\kappa^{2} (n-1)^{2}}(\phi+1)^{-2},
\end{equation}
and 
\begin{equation}
\eta(\phi)=\frac{n(2n-1)}{\kappa^{2} (n-1)^{2}}(\phi+1)^{-2},
\end{equation}
respectively.
The spectral index defined in Eq. \eqref{ns} has the form
\begin{equation}
n_{s}=1+\frac{(-8n^{2}+10n-3)}{\kappa^{2}(n^{2}-2n+1)}(\phi+1)^{-2}.\label{nsmodel1}
\end{equation}
The tensor-to-scalar ratio $r$ defined in Eq. \eqref{r} becomes
\begin{equation}
r=\frac{8}{\kappa^{2}}\frac{(2n-1)^{2}}{(n-1)^{2}}(\phi+1)^{-2}.\label{rmodel1}
\end{equation} 
The computations of numerical values of these expressions of $n_{s}$ and $r$ obtained in Equations $(\ref{nsmodel1})$ and $(\ref{rmodel1})$
are presented in table $\ref{Tablemodel1}$.
Let us onsider Eq. \eqref{kkg1} and use the condition of the slow-roll presented in Eq. \eqref{slow2}, i.e., the limiting case of the scalar field $\phi$ slowly evolving over time. When we apply this assumption to Eq. \eqref{kkg1} we have,
\begin{equation}
-\frac{1}{3}\Big[\frac{\beta(2-n)}{(n\beta)^{n/(n-1)}}(\phi+1)^{n/(n-1)}+\frac{3w-1}{\phi+1}\tilde{\mu}_{m}\Big]<V'(\phi).
\end{equation}
Using the expression derived  above about $V'(\phi)$, we have the condition that
\begin{equation}
\tilde{\mu}_{m}<\frac{2\beta(2-n)}{(3w-1)(n\beta)^{n/(n-1)}}(\phi+1)^{n/(n-1)}.\label{mcondition1}
\end{equation}
From this inequality, one can see that from the slow-roll approximation, the scalar field $\phi$ is part of the expression
that puts the condition on the matter energy density $\tilde{\mu}_{m}$. 

 \subsection{Slow-roll approximation of $f(R)=\alpha R+\beta R^{n}$}

For the case of $f(R)=\alpha R+\beta R^{n}$, the slow-roll parameters $\epsilon(\phi)$ and $\eta(\phi)$ are given as
\begin{equation}
\epsilon(\phi)=\frac{\big(2\alpha (n-1)-(n-2)(\phi+1)\big)^{2}(\phi+1-\alpha)^{2}}{6\kappa^{2}n^{4}\beta^{2}(n-1)^{2}(3\alpha(n-1)-(n-2)(\phi+1))^{2}},
\end{equation}
and
\begin{equation}
\eta(\phi)=\frac{\beta(2-n)[\frac{1}{\phi+1-\alpha}+\frac{2n\alpha}{2-n}+\frac{n(\phi+1)}{n-1}]}{3\kappa^{2}(1-n)(\alpha-\phi-1)[n(3\alpha-\phi-1)-3\alpha+2(\phi+1)]},
\end{equation}
respectively.
The spectral index defined in Eq. \eqref{ns} becomes
\begin{equation*}
n_{s}=1-\frac{\Big(2\alpha(n-1)-(n-2)(\phi+1)\Big)^{2}(\phi+1-\alpha)^{2}}{\kappa^{2}(n-1)^{2}n^{4}[3\alpha(n-1)-(n-2)(\phi+1)]^{2}}+
\end{equation*}
\begin{equation}
\frac{2(2-n)\beta[(\phi+1-\alpha)^{-1}+\frac{2n\alpha}{(2-n)}+\frac{n(\phi+1)}{(n-1)}]}{
3\kappa^{2}(1-n)(-\alpha-\phi-1)[n(3\alpha-\phi-1)-3\alpha+2(\phi+1)]}.\label{nsmodel2}
\end{equation}
The tensor-to-scalar ratio $r$ defined in Eq. \eqref{r} becomes
\begin{equation}
r=\frac{8\Big(2\alpha(n-1)-(n-2)(\phi+1)\Big)^{2}(\phi+1-\alpha)^{2}}{3\kappa^{2}(n-1)^{2}n^{4}[3\alpha(n-1)-(n-2)(\phi+1)]^{2}}.\label{rmodel2}
\end{equation}
The computations of numerical values of these expressions of $n_{s}$ and $r$ obtained in Equations $(\ref{nsmodel2})$ and $(\ref{rmodel2})$
are presented in table $\ref{Tablemodel2}$.

 \subsection{Slow-roll approximation for $f(R)=R-\frac{\nu^{4}}{R}$ model }

For the case of $f(R)=R-\frac{\nu^{4}}{R}$, with consideration of the first root in Ricci scalar, the slow-roll parameters $\epsilon(\phi)$ and $\eta(\phi)$ are given as
\begin{equation}
\epsilon(\phi)=\frac{1}{2\kappa^{2}}\Big[\frac{15(\phi^{1/2}-\phi^{3/2}-2\nu^{8}\phi^{-1/2})}{10\phi^{3/2}-6\phi^{5/2}-
60\nu^{8}\phi^{1/2}}\Big]^{2}\;,
\end{equation}
and
\begin{equation}
\eta(\phi)=\frac{1}{\kappa^{2}}\Big(\frac{15\phi^{-1/2}-30\nu^{8}\phi^{-3/2}-45\phi^{1/2}}{20\phi^{3/2}-12\phi^{5/2}-
120\nu^{8}\phi^{1/2}}\Big)
\end{equation}
respectively. The spectral index defined in Eq. \eqref{ns} becomes
\begin{equation*}
n_{s}=1-\frac{3}{\kappa^{2}}\Big[\frac{15(\phi^{1/2}-\phi^{3/2}-2\nu^{8}\phi^{-1/2})}{10\phi^{3/2}-6\phi^{5/2}-
60\nu^{8}\phi^{1/2}}\Big]^{2}+
\end{equation*}
\begin{equation}
\frac{2}{\kappa^{2}}\Big(\frac{15\phi^{-1/2}-30\nu^{8}\phi^{-3/2}-45\phi^{1/2}}{20\phi^{3/2}-12\phi^{5/2}-
120\nu^{8}\phi^{1/2}}\Big).\label{nsmodel3a}
\end{equation}
The tensor-to-scalar ratio $r$ defined in Eq. \eqref{r} becomes
\begin{equation}
r=\frac{8}{\kappa^{2}}\Big[\frac{15(\phi^{1/2}-\phi^{3/2}-2\nu^{8}\phi^{-1/2})}{10\phi^{3/2}-6\phi^{5/2}-
60\nu^{8}\phi^{1/2}}\Big]^{2}.\label{rmodel3a}
\end{equation}
The computations of numerical values of these expressions of $n_{s}$ and $r$ obtained in Equations $(\ref{nsmodel3a})$ and $(\ref{rmodel3a})$
for root $R_{+}$ are presented in table $\ref{Tablemodel3a}$.
When one considers the other (negative) root in the Ricci scalar, the slow-roll parameters $\epsilon(\phi)$ and $\eta(\phi)$ are given as
\begin{equation}
\epsilon(\phi)=\frac{1}{2\kappa^{2}}\Big[\frac{-15\phi^{1/2}+15\phi^{3/2}+30\nu^{8}\phi^{-1/2})}{-10\phi^{3/2}+6\phi^{5/2}+
60\nu^{8}\phi^{1/2}}\Big]^{2}\;,
\end{equation}
and
\begin{equation}
\eta(\phi)=\frac{1}{\kappa^{2}}\Big(\frac{-15\phi^{-1/2}-30\nu^{8}\phi^{-3/2}+45\phi^{1/2}}{-20\phi^{3/2}+12\phi^{5/2}+
120\nu^{8}\phi^{1/2}}\Big)
\end{equation}
respectively. The spectral index defined in Eq. \eqref{ns} becomes
\begin{equation*}
n_{s}=1-\frac{3}{\kappa^{2}}\Big[\frac{-15\phi^{1/2}+15\phi^{3/2}+30\nu^{8}\phi^{-1/2})}{-10\phi^{3/2}+6\phi^{5/2}+
60\nu^{8}\phi^{1/2}}\Big]^{2}+
\end{equation*}
\begin{equation}
\frac{2}{\kappa^{2}}\Big(\frac{-15\phi^{-1/2}-30\nu^{8}\phi^{-3/2}+45\phi^{1/2}}{-20\phi^{3/2}+12\phi^{5/2}+
120\nu^{8}\phi^{1/2}}\Big).\label{nsmodel3b}
\end{equation}
The tensor-to-scalar ratio $r$ defined in Eq. \eqref{r} becomes
\begin{equation}
r=\frac{8}{\kappa^{2}}\Big[\frac{-15\phi^{1/2}+15\phi^{3/2}+30\nu^{8}\phi^{-1/2})}{-10\phi^{3/2}+6\phi^{5/2}+
60\nu^{8}\phi^{1/2}}\Big]^{2}.\label{rmodel3b}
\end{equation}
The computations of numerical values of these expressions of $n_{s}$ and $r$ obtained in Equations $(\ref{nsmodel3b})$ and $(\ref{rmodel3b})$
for root $R_{-}$ are presented in table $\ref{Tablemodel3b}$.
 \subsection{Slow-roll approximation for \\
 $f(R)=R-(1-m)\nu^{2}\Big(\frac{R}{\nu^{2}}\Big)^{m}-2\Lambda$ model }

For the case of $f(R)=R-(1-m)\nu^{2}\Big(\frac{R}{\nu^{2}}\Big)^{m}-2\Lambda$, the slow-roll parameters $\epsilon(\phi)$ and $\eta(\phi)$ are given as
\begin{equation}
\epsilon(\phi)=\frac{1}{2\kappa^{2}}\left[\frac{M_{1}-M_{2}-M_{3}-4\Lambda}{M_{4}+M_{5}-M_{6}-4\Lambda\phi}\right]^{2},
\end{equation}
and
\begin{equation}
\eta(\phi)=\frac{1}{\kappa^{2}}\left(\frac{M_{7}+M_{8}-M_{9}}{M_{4}+M_{5}-M_{6}-4\Lambda\phi}\right),
\end{equation}
where $M_{1}=\frac{\nu^{2}\phi^{\frac{1}{(m-1)}}}{[m(m-1)]^{\frac{1}{(m-1)}}}$, 
$M_{2}=\frac{2(1-m)\nu^{2}\phi^{\frac{m}{(m-1)}}}{[m(m-1)]^{\frac{m}{(m-1)}}},$\\
$ M_{3}=\frac{\nu^{2}\phi^{\frac{m}{(m-1)}}}{[m(m-1)]^{\frac{1}{(m-1)}}}$,
$M_{4}=\frac{(m-1)\nu^{2}\phi^{\frac{m}{(m-1)}}}{m[m(m-1)]^{\frac{1}{(m-1)}}}$,\\
$M_{5}=\frac{2(m-1)^{2}\nu^{2}\phi^{\frac{(2m-1)}{(m-1)}}}{(2m-1)[m(m-1)]^{\frac{m}{(m-1)}}}$,
$M_{6}=\frac{(m-1)\nu^{2}\phi^{\frac{(2m-1)}{(m-1)}}}{(2m-1)[m(m-1)]^{\frac{1}{(m-1)}}}$,\\
$M_{7}=\frac{m\nu^{2}\phi^{(2-m)/(m-1)}}{(m-1)[m(m-1)]^{1/(m-1)}}$, $M_{8}=\frac{2\nu^{2}\phi^{1/(m-1)}}{m[m(m-1)]^{m/(m-1)}}$,\\
$M_{9}=\frac{m\nu^{2}\phi^{m/(m-1)}}{(m-1)[m(m-1)]^{1/(m-1)}}$
respectively. The spectral index defined in Eq. \eqref{ns} becomes
\begin{equation}
\begin{split}
 n_{s}=&1-\frac{3}{\kappa^{2}}\left[\frac{M_{1}-M_{2}-M_{3}-4\Lambda}{M_{4}+M_{5}-M_{6}-4\Lambda\phi}\right]^{2}+\\
       &+\frac{2}{\kappa^{2}}\left(\frac{M_{7}+M_{8}-M_{9}}{M_{4}+M_{5}-M_{6}-4\Lambda\phi}\right).\label{nsmodel4}
\end{split}
\end{equation}
The tensor-to-scalar ratio $r$ defined in Eq. \eqref{r} becomes
\begin{equation}
r=\frac{8}{\kappa^{2}}\left[\frac{M_{1}-M_{2}-M_{3}-4\Lambda}{M_{4}+M_{5}-M_{6}-4\Lambda\phi}\right]^{2}.\label{rmodel4}
\end{equation}
The computations of numerical values of these expressions of $n_{s}$ and $r$ obtained in Equations $(\ref{nsmodel4})$ and $(\ref{rmodel4})$
are presented in table $\ref{Tablemodel4}$.

From different expressions obtained for spectral index $n_{s}$, tensor-to-scalar ratio $r$ during the slow-roll approximations for the 
four considered $f(R)$ models, one can see that the scalar field $\phi$ is playing a major role in controlling the bahaviour in the respective
expressions more than the other parameters. One can constrain the $f(R)$ models considered in this work by using the observed spectral
index values, for more details on approximate values of $n_{s}$, for example, see Ref. \cite{Spergel1}. The observational value of 
the tensor-to-scalar ratio $r$ can also constrain the $f(R)$ models, for current observed values, see,  for example,  Ref. \cite{Spergel2}.
The result from full Planck Survey (see Ref. \cite{ade2016planck}) shows that spectral index of curvature perturbations $n_{s}=0.968\pm 0.006$. 
It also provides the upper bound on the tensor-to-scalar ratio to be $r<0.11 (95\% CL)$, which is consistent with the result obtained
in the joint analysis done from BICEP2/Keck Array and Planck data (see Ref. \cite{ade2015joint}) where $r<0.2 (95\% CL)$. We computed numerically 
spectral index $n_{s}$ and tensor-to-scalar ratio $r$ for the four $f(R)$ models considered in this work. One can see that the results 
from both Planck survey and BICEP2/Keck Array puts constraints on the values of the power of $f(R)$ models under consideration. 
The numerical results obtained in this work is in agreement with those obtained in Ref. \cite{bamba2014inflationary} 
where deep analysis of slow roll inflation has been done with perfect fluid consideration. 

For $f(R)=\beta R^{n}$ model, numerical values for spectral index $n_{s}$ presented in Equation $(\ref{nsmodel1})$ and 
that of tensor-to-scalar ratio $r$ presented in Equation $(\ref{rmodel1})$ are listed
in Table \ref{Tablemodel1}. One can notice that the approximate values of $r$ are very close to that of Planck survey but fot $n_{s}$ 
the values tend to be close to unit yet the Planck survey's value is slightly less than unit.

For $f(R)=\alpha R+\beta R^{n}$ model, numerical values for spectral index $n_{s}$ presented in Equation $(\ref{nsmodel2})$ and 
that of tensor-to-scalar ratio $r$ presented in Equation $(\ref{rmodel2})$ are listed
in Table \ref{Tablemodel2}. It this table, we can see that for specific values of $\alpha,\beta,n$ amd $\phi$, 
we have obtained the values of $n_{s}$ and $r$ close to the Planck survey results.

For $f(R)=R-\frac{\nu^{4}}{R}$ model, with positive root $(R_{+})$, numerical values for spectral index $n_{s}$ 
presented in Equation $(\ref{nsmodel3a})$ and 
that of tensor-to-scalar ratio $r$ presented in Equation $(\ref{rmodel3a})$ are listed
in Table \ref{Tablemodel3a} for positive root $R_{+}$. It this table, we can see that the values of $n_{s}$ and $r$
are slightly close to the Planck survey results. For negative root $(R_{-})$, numerical values for spectral index $n_{s}$ presented in Equation $(\ref{nsmodel3b})$ and 
that of tensor-to-scalar ratio $r$ presented in Equation $(\ref{rmodel3b})$ are listed
in Table \ref{Tablemodel3b} for negative root $R_{-}$. It this table, we have obtained the values of $n_{s}$ and $r$ that fit under predictions of
Planck survey results. 

For $f(R)=R-(1-m)\nu^{2}\Big(\frac{R}{\nu^{2}}\Big)^{m}-2\Lambda$ model, numerical values for spectral index $n_{s}$ presented in Equation $(\ref{nsmodel4})$ and 
that of tensor-to-scalar ratio $r$ presented in Equation $(\ref{rmodel4})$ are listed
in Table \ref{Tablemodel4}. It this table, we have obtained the values of $n_{s}$ and $r$ very close to the Planck survey results. 

\begin{table}[!h]
  \begin{tabular}{|l | l |c |r|r|r|  }
    \hline
   n& $\phi$ & r & $n_{s}$ & r (Planck data)& $n_{s}$ (Planck data)  \\ \hline
    1.99 & 24 & 0.1159 & 1.0009 & $<$0.11 (95 $\%$ CL)&0.968$\pm$ 0.006    \\ \hline 
     1.99 & 25& 0.1072 & 1.0017&$<$0.11 (95 $\%$ CL)&0.968$\pm$ 0.006 \\ \hline
     1.99& 26 & 0.0999 & 1.0016&$<$0.11 (95 $\%$ CL)&0.968$\pm$ 0.006 \\ \hline 
     1.99& 30 & 0.075 & 1.0012 &$<$0.11 (95 $\%$ CL)&0.968$\pm$ 0.006\\ \hline 
     
     1.5& 32 & 0.1175 & 1.0017 &$<$0.11 (95 $\%$ CL)&0.968$\pm$ 0.006\\ \hline 
     1.5& 33& 0.1072 & 1.0010&$<$0.11 (95 $\%$ CL)&0.968$\pm$ 0.006 \\ \hline
     1.5& 34 & 0.1044 & 1.0015&$<$0.11 (95 $\%$ CL)&0.968$\pm$ 0.006 \\ \hline 
     1.5& 35 & 0.098 & 1.0014 &$<$0.11 (95 $\%$ CL)&0.968$\pm$ 0.006\\ \hline 
     
     1.2 & 57 & 0.1165 & 1.0014&$<$0.11 (95 $\%$ CL)&0.968$\pm$ 0.006 \\ \hline 
     1.2 & 58& 0.1126 & 1.0014&$<$0.11 (95 $\%$ CL)&0.968$\pm$ 0.006\\ \hline
     1.2& 59&  0.1088 & 1.0013&$<$0.11 (95 $\%$ CL)&0.968$\pm$ 0.006 \\ \hline 
     1.2& 60 & 0.1053 & 1.0013&$<$0.11 (95 $\%$ CL)&0.968$\pm$ 0.006 \\ \hline 
  
  \end{tabular}
 \caption{Table about numerical values of spectral index $n_{s}$ and tensor-to-scalar ratio $r$ with the Planck survey data for easy comparison,
  for $f(R)=\beta R^{n}$.}  \label{Tablemodel1}
  \end{table}

\begin{table}[!h]
  \begin{tabular}{|l | l |c |r|r|r|r|r|  }
    \hline
   n& $\phi$&$\alpha$ & $\beta$ &r & $n_{s}$& r (Planck data)& $n_{s}$ (Planck data)    \\ \hline
    1.5 & 1.1& 0.002 & 0.001 & 0.0586 & 0.97800&$<$0.11 (95 $\%$ CL)&0.968$\pm$ 0.006 \\ \hline 
     1.5& 1.2& 0.002 & 0.001 & 0.0644 & 0.97585 &$<$0.11 (95 $\%$ CL)&0.968$\pm$ 0.006\\ \hline
     1.5& 1.3& 0.002 & 0.001 & 0.0704 & 0.97360 &$<$0.11 (95 $\%$ CL)&0.968$\pm$ 0.006\\ \hline 
     1.5& 1.4&0.002 & 0.001 &  0.0766&  0.97125 &$<$0.11 (95 $\%$ CL)&0.968$\pm$ 0.006\\ \hline   
     1.5& 1.5& 0.002 & 0.001 & 0.0832&  0.96880 &$<$0.11 (95 $\%$ CL)&0.968$\pm$ 0.006\\ \hline 
     1.5& 1.6& 0.002 & 0.001 & 0.0900 & 0.96625  &$<$0.11 (95 $\%$ CL)&0.968$\pm$ 0.006\\ \hline
     1.5& 1.7& 0.002 & 0.001 & 0.0970&  0.96360 &$<$0.11 (95 $\%$ CL)&0.968$\pm$ 0.006 \\ \hline 
     1.5& 1.8& 0.002 & 0.001 & 0.1044&  0.96085 &$<$0.11 (95 $\%$ CL)&0.968$\pm$ 0.006\\ \hline 
     1.5& 1.9& 0.002 & 0.001 & 0.1120&  0.95800 &$<$0.11 (95 $\%$ CL)&0.968$\pm$ 0.006\\ \hline 
    
  \end{tabular}
  \caption{Table about numerical values of spectral index $n_{s}$ and tensor-to-scalar ratio $r$ with the Planck survey data for easy comparison,
  for $f(R)=\alpha R+\beta R^{n}$.} \label{Tablemodel2}
  \end{table}
  
\begin{table}[!h]
 \begin{tabular}{|l | l |c |r|r|r|r|r|  }
    \hline
   $\nu$& $\phi$ &r & $n_{s}$ & r (Planck data)& $n_{s}$ (Planck data)   \\ \hline
    0.5 & 0.3 & 0.1897 & 0.96113&$<$0.11 (95 $\%$ CL)&0.968$\pm$ 0.006 \\ \hline 
    0.5&  0.301 & 0.18809 & 0.96033 &$<$0.11 (95 $\%$ CL)&0.968$\pm$ 0.006 \\ \hline
     0.5& 0.302 & 0.18642 & 0.95954 &$<$0.11 (95 $\%$ CL)&0.968$\pm$ 0.006\\ \hline 
     0.5& 0.36 &  0.1139&  0.92319 &$<$0.11 (95 $\%$ CL)&0.968$\pm$ 0.006\\ \hline   
     0.5&  0.365 & 0.1094&  0.9206 &$<$0.11 (95 $\%$ CL)&0.968$\pm$ 0.006\\ \hline 
     0.5& 0.369 & 0.1059 & 0.91865  &$<$0.11 (95 $\%$ CL)&0.968$\pm$ 0.006\\ \hline
     
     0.6& 2.0  & 0.052 & 0.99621 &$<$0.11 (95 $\%$ CL)&0.968$\pm$ 0.006   \\ \hline 
     0.6& 2.1  & 0.0672 & 0.98895  &$<$0.11 (95 $\%$ CL)&0.968$\pm$ 0.006 \\ \hline 
     0.6& 2.2  & 0.0865 & 0.98027  &$<$0.11 (95 $\%$ CL)&0.968$\pm$ 0.006\\ \hline 
     0.6& 2.29 & 0.1093 & 0.97061   &$<$0.11 (95 $\%$ CL)&0.968$\pm$ 0.006 \\ \hline 
     0.6& 2.299& 0.11198& 0.96952 &$<$0.11 (95 $\%$ CL)&0.968$\pm$ 0.006  \\ \hline 
     0.6& 2.3  & 0.1122 & 0.96940 &$<$0.11 (95 $\%$ CL)&0.968$\pm$ 0.006   \\ \hline 
     0.6& 2.365& 0.1338 & 0.96059 &$<$0.11 (95 $\%$ CL)&0.968$\pm$ 0.006 \\ \hline 
 
  \end{tabular}
  \caption{Table about numerical values of spectral index $n_{s}$ and tensor-to-scalar ratio $r$ with the Planck survey data for easy comparison,
  for $f(R)=R+\frac{\nu^{4}}{R}$ with positive root $(R_{+})$.}\label{Tablemodel3a}
\end{table}
 
\begin{table}[!h]
\begin{tabular}{|l | l |c |r|r|r|r|r|  }
    \hline
   $\nu$& $\phi$ &r & $n_{s}$& r (Planck data)& $n_{s}$ (Planck data)    \\ \hline
    0.9 & 0.35 & 0.0702 & 0.94526 &$<$0.11 (95 $\%$ CL)&0.968$\pm$ 0.006\\ \hline 
    0.9&  0.40 & 0.052 & 0.95977 &$<$0.11 (95 $\%$ CL)&0.968$\pm$ 0.006 \\ \hline
     0.9& 0.401 & 0.0526 & 0.960014 &$<$0.11 (95 $\%$ CL)&0.968$\pm$ 0.006\\ \hline 
     0.9& 0.402&  0.0524&  0.960248&$<$0.11 (95 $\%$ CL)&0.968$\pm$ 0.006 \\ \hline   
     0.9&  0.403 & 0.052134&  0.960480&$<$0.11 (95 $\%$ CL)&0.968$\pm$ 0.006 \\ \hline 
     0.9& 0.405& 0.0516 & 0.960940  &$<$0.11 (95 $\%$ CL)&0.968$\pm$ 0.006\\ \hline

  \end{tabular}
   \caption{Table about numerical values of spectral index $n_{s}$ and tensor-to-scalar ratio $r$ with the Planck survey data for easy comparison,
  for $f(R)=R+\frac{\nu^{4}}{R}$ with negative root $(R_{-})$.}\label{Tablemodel3b}
\end{table}  

\begin{table}[!h]
 \begin{tabular}{|l | l |c |r|r|r|r|r|r|  }
    \hline
   $m$&$\nu$& $\phi$ &r & $n_{s}$ & r(Planck data)& $n_{s}$ (Planck data)   \\ \hline
    1.5&1 & 7.5 & 0.01202 & 0.959644 &$<$0.11 (95 $\%$ CL)&0.968$\pm$ 0.006\\ \hline 
    1.5&1&  7.55 & 0.01187 & 0.959825 &$<$0.11 (95 $\%$ CL)&0.968$\pm$ 0.006\\ \hline
    1.5&1& 7.6 & 0.01173 & 0.960004&$<$0.11 (95 $\%$ CL)&0.968$\pm$ 0.006 \\ \hline 
    1.5&1& 7.65&  0.01158&  0.960182 &$<$0.11 (95 $\%$ CL)&0.968$\pm$ 0.006\\ \hline   
    1.5&1&  7.7 & 0.01144&  0.960358 &$<$0.11 (95 $\%$ CL)&0.968$\pm$ 0.006 \\ \hline 
    1.55&1& 7.0& 0.0120 & 0.959103 &$<$0.11 (95 $\%$ CL)&0.968$\pm$ 0.006 \\ \hline 
    1.55&1& 7.2&  0.01147&  0.959817&$<$0.11 (95 $\%$ CL)&0.968$\pm$ 0.006 \\ \hline   
    1.55&1& 7.3 & 0.0118&  0.9601652 &$<$0.11 (95 $\%$ CL)&0.968$\pm$ 0.006\\ \hline 
    1.55&1& 7.4& 0.0109 & 0.9605071 &$<$0.11 (95 $\%$ CL)&0.968$\pm$ 0.006 \\ \hline
    
  \end{tabular}
\caption{Table about numerical values of spectral index $n_{s}$ and tensor-to-scalar ratio $r$ with the Planck survey data for easy comparison, for 
    $f(R)=R-(1-m)\nu^{2}\left(\frac{R}{\nu^{2}}\right)^{m}-2\Lambda$.} \label{Tablemodel4} 
  \end{table}

  \section{Conclusion}

We considered $f(R)$ gravity in scalar-tensor language and we reviewed the thermodynamic quantities of the FRLW. 
The scalar-tensor language of $f(R)$ gravity has been revisited and it has been shown that $f(R)$ gravity is a sub-class of
Brans-Dicke scalar-tensor theory of gravity with the coupling constant $\omega=0$. In that line, four $f(R)$ models have been considered.
It is found in this work that for the first model, the scalar potential $V(\phi)$ depends on scalar field $\phi$ in a parabolic way. 
For the $n=1.99$ and more generally for $n<2$, the potential has the positive concavity and this is in agreement with the existing 
literature about the inflation potential. When $n>2$, the potential decreases, thus showing instability. The same inspections are applied
to the second model. After stressing scalar field $\phi$ to depend on time and assuming that at the early universe the scalar field 
was dominating the matter, we have obtained the solution to Klein-Gordon equation for both models. It was found for the first model that, 
as the time grows the scalar field decreases and approaches zero asymptotically. This inspection is also in agreement with the 
literature on the scalar field behavior. The same approach has been applied to the second model and it was found that as we stress
$n$ to be constant and keep $\beta$ changing while still keeping $\alpha$ constant, the function $\phi(t)$ crosses the $t$-axis at 
different values. 

For the third model, one might see that like in the case of the positive root (see Fig. $\ref{pot3a}$), the same inspection can be drawn that as the 
value of $\nu$ is small the potential $V(\phi)$ is becoming the one which supports the inflation.

For the fourth $f(R)$ model, we have obtained that there is a transition between $1.5<m<1.55$. The behavior of the potentials
with $m<1.5$ is totally different from those with $m>1.55$. Therefore the potential in $U$-form is only supported for the 
potential $V(\phi)$ with $m<1.55$. Thus, it can be concluded that the inflation potential shape puts a constraint on this $f(R)$ model.

The slow-roll approximation has been considered for the four f(R) models and we have obtained the respective expressions for spectral 
index $n_{s}$ and the tensor-to-scalar ratio $r$ explicitly. We have obtained numerical values which are very close to the ones from 
Planck data. For example, for the second model we have generated $n_{s}=0.96085$ and $r=0.1044$. For the third model we got 
$n_{s}=0.96940$ and $r=0.1122$ with the root $R_{+}$ and $n_{s}=0.960014$ and $r=0.0526$ with root $R_{-}$. For the fourth model
we generated $n_{s}=0.960182$ and $r=0.0115$. This shows that if precise values of these two parameters are  measured from current
and upcoming observational probes, we can, in principle, constrain the viability of the different $f(R)$ models considered in this preliminary work. More rigorous calculations, both exact and computational, under more realistic assumptions  and for more realistic toy models is left for future work.


\section*{Acknowledgments} 

JN gratefully acknowledges financial support from the Swedish International Development Cooperation Agency (SIDA) through the International Science Program (ISP) 
to the University of Rwanda (Rwanda Astrophysics, Space and Climate Science Research Group), and the Entoto Observatory and Research Center  for partial support.  

\bibliographystyle{unsrt}
\bibliography{references}

\end{document}